# Communicability and Multipartite Structures in Complex Networks at Negative Absolute Temperatures

Ernesto Estrada[1,3*], Desmond J. Higham[2] and Naomichi Hatano[3]

[1]Institute of Complexity Science, Department of Physics and Department of Mathematics, University of Strathclyde, Glasgow G1 1XH, UK

[2]Department of Mathematics, University of Strathclyde, Glasgow G1 1XH, UK

[3]Institute of Industrial Science, University of Tokyo, Komaba, Meguro, 153-8505, Japan

[*] Corresponding author. E-mail: estrada66@yahoo.com

**Abstract**

We here present a method of clearly identifying multi-partite subgraphs in a network. The method is based on a recently introduced concept of the communicability, which very clearly identifies communities in a complex network. We here show that, while the communicability at a positive temperature is useful in identifying communities, the communicability at a negative temperature is useful in idenfitying multi-partitite subgraphs; the latter quantity between two nodes is positive when the two nodes belong to the same subgraph and is negative when not. The method is able to discover `almost' multi-partite structures, where inter-community connections vastly outweigh intra-community connections. We illustrate the relevance of this work to real-life food web and protein-protein interaction networks.

**Introduction**

Since the publication of the seminal paper by Girvan and Newman in 2002 [1], there has been a dramatic explosion of strategies and approaches for detecting communities in complex networks [2-8]. These networks represent systems in which the nodes account for the interacting entities, such as molecules, cells, animal species, technological artifacts, etc., and the links record their interactions. A network community is a group of entities which have a large number of interactions among them but very few iteractions with members of the other groups [1].

A completely contrasting type of structure can also be found in complex networks. In such structures the nodes are organized in groups of (almost) bipartite subgraphs [9, 10]. Here, the phrase bipartite signifies that the (sub)graph can be divided into two distinct subsets such that all connections are from one subset to the other. In this organization a community is a collection of nodes that are not interconnected, but have many neighbors in common. These structures appear in a wide variety of complex systems. For instance, in the context of protein-protein interaction networks, the bipartite structure is intimately related to the existence of complementary binding domains in proteins [11, 12] as well as to the identification of essential proteins [13]. In food webs, bipartite substructures might represent different trophic levels, such as those formed by plants and hervibores, and in social networks some of these disjoint sets can represent potential competitors as in sellers-buyers or dating networks.

The main purpose of the present paper is to identify a bipartite structure (and more generally a multipartitite structure) in an arbitrary network. To do this, we introduce the concept of communicability with a negative temperature. Two of the authors recently demonstrated that we can identify communities in an arbitrary network very clearly with the use of the communicability (with a positive temperature) [14]. Here we show that we can clearly identify a multipartite structure with the use of the communicability with a negative temperature.

**Preliminaries**

We represent a complex network by an undirected graph $G = (V, E)$, where $V$ and $E$ are the sets of nodes and links, respectively. Let $G$ have $N$ nodes. Then the adjacency matrix of $G$, $\mathbf{A}(G) = \mathbf{A}$, is a square, symmetric matrix of order $N$, whose elements $A_{ij}$ are ones or zeroes if the corresponding nodes are adjacent or not, respectively. This matrix has $N$ (not necessarily distinct) real-valued eigenvalues [15], which are denoted here by $\lambda_1, \lambda_2, \ldots, \lambda_N$, and assumed to be labelled in a non-increasing manner: $\lambda_1 \geq \lambda_2 \geq \ldots \geq \lambda_N$. Let $\tilde{\mathbf{a}}_j$ be an orthonormal eigenvector corresponding to the eigenvalue $\lambda_j$. Then, $\gamma_j(i)$ designates the component of this eigenvector to the $i$th node in the network. A graph is said to be bipartite if its nodes can be divided into two disjoint sets $V_1$ and $V_2$ such that every link connects a vertex in $V_1$ and one in $V_2$, but there is no edge between two nodes in the same set.

**Theoretical Approach**

Let us consider the communicability between a pair of nodes $p$ and $q$ in the network [14],

$$G_{pq}(\beta) = \left(e^{\beta \mathbf{A}}\right)_{pq} = \sum_{j=1}^{n} \phi_j(p) \phi_j(q) e^{\beta \lambda_j}, \tag{1}$$

where $\beta = \dfrac{1}{kT}$ is the inverse temperature, $k$ is the Boltzman constant and $T$ is the absolute temperature [16]. We have previously shown that (1) represents the Green's function of the network, that is, a function which expresses how an impact propagates from one node to another node in the graph [14]. Using the quantity (1), we constructed the "communicability graph $\Theta(G)$" from the original graph $G$. The communicability graph indicates communities in the graph $G$ very clearly [17].

It is known from spectral clustering techniques that the eigenvectors corresponding to positive eigenvalues give a partition of the network into clusters of tightly connected nodes [18, 19]. In contrast, the eigenvectors corresponding to negative eigenvalues make partitions in which nodes are not close to those which they are linked, but rather with those with which they are not linked [18, 19]. Such differences have made possible the classification of complex networks into four universal classes [20]. Let us demonstrate the above statements for a cycle $C_n$ with even number of nodes $n$. The adjacency matrix of a cycle is diagonalized by the eigenvectors $\{\phi_j(p) = \operatorname{Re} e^{ik_j p}/\sqrt{n}\}$, where $\phi_j(p)$ denotes the component on the $p$ th node of the eigenvector with the label $j$, and $k_j \equiv 2\pi(j-1)/n$. The corresponding eigenvalues are $\lambda_j = 2\cos k_j$. The largest eigenvalue $\lambda_1 = 2$ is given by the eigenvector $\phi_1(p) = \text{const.}$, which is a partition of the whole network $C_n$ into just one cluster. The second largest eigenvalue $\lambda_2 = 2\cos(2\pi/n)$ is given by the eigenvector $\phi_2(p) = \cos(2\pi p/n)/\sqrt{n}$, which is positive for almost half of the nodes and negative for the other half. In short, the second largest eigenvalue gives a partition of the network $C_n$ into two clusters. On the other hand, the lowest eigenvalue $\lambda_{1+n/2} = -2$ is given by the eigenvector $\phi_{1+n/2}(p) = (-1)^p/\sqrt{n}$, which gives a partition of the network $C_n$ into two subgraphs; that is, the eigenvetor is positive for the nodes with even $p$ and negative for the nodes with odd $p$.

From the perspective of the communicability function (1) we can say that a positive (negative) value of $\beta$ increases the contribution of the positive (negative) eigenvalues to the communicability function. Then if we write the communicability function as

$$G_{pq}(\beta) = \sum_{\lambda_j<0} \phi_j(p)\phi_j(q)e^{\beta\lambda_j} + \sum_{\lambda_j=0} \phi_j(p)\phi_j(q)e^{\beta\lambda_j} + \sum_{\lambda_j>0} \phi_j(p)\phi_j(q)e^{\beta\lambda_j} \qquad (2)$$

we have that

$$G_{pq}(\beta>0) \approx \sum_{\lambda_j>0}^{n} \phi_j(p)\phi_j(q)e^{\beta\lambda_j} \qquad (3)$$

$$G_{pq}(\beta<0) \approx \sum_{\lambda_j<0}^{n} \phi_j(p)\phi_j(q)e^{-|\beta|\lambda_j} \qquad (4)$$

In other words, $G_{pq}(\beta>0)$ determines a partition of the network into clusters of tightly connected nodes, which corresponds to the network communities. On the other hand, for $G_{pq}(\beta<0)$ the network is partitioned in such a way that the nodes are close to other nodes which have similar patterns of connections with other sets of nodes, i.e., nodes to which they are structurally equivalent. In the first case, we say that the nodes corresponding to larger components tend to form *quasi-cliques*. That is, clusters in which every two nodes tend to interact with each other. In the second case, the nodes tend to form *quasi-bipartites*, i.e., nodes are partitioned into *almost* disjoint subsets with high connectivity between sets but low internal connectivity.

Negative values of $\beta$ arise when the absolute temperature is negative $(T<0)$. Note that the temperature scale from cold to hot then runs [21] $0K,\cdots,+300K,\cdots,+\infty K,\cdots,-\infty K,\cdots,-300K,\cdots,-0K$. In the limit $T \to -0$ the largest contribution to the communicability is from the lowest eigenvalue of the adjacency matrix, $\lambda_n$

$$\lim_{\beta \to -\infty} G_{pq}(\beta) \approx \phi_n(p)\phi_n(q)e^{-|\beta|\lambda_n} \qquad (5)$$

which is known to produce a two-coloring of the nodes [22]. In the above-mentioned example of the cycle $C_n$, the quantity (5) is positive when $p$ and $q$ have the same parity (that is, when they belong to the same subgraph) and negative when not. This implies that the sign of the communicability at a negative temperature indicates whether two nodes belong to the same subgraph or not. This is the main observation on which we develop the theoretical approach hereafter.

In order to understand the meaning of the inverse temperature $\beta$ in the context of complex networks, we may express the communicability in terms of powers of the adjacency matrix,

$$\left(e^{\beta A}\right)_{pq} = \sum_{k=0}^{\infty} \frac{\beta^k \left(A^k\right)_{pq}}{k!} \qquad (11)$$

Accordingly, $\beta$ represents a weight given to every link of the network. This weight accounts for the "strength" of the interaction between the corresponding nodes in the graph. For instance, $\beta = 0$, which corresponds to the limit $T \to \infty$, corresponds to a graph with no links. This case is similar to a gas formed by monoatomic particles. On the other hand, very large values of $\beta$ in the limit $T \to +0$ represents very large attractive interactions between pairs of bonded nodes in a similar manner to a solid. The new cases considered in this work, $\beta < 0$, correspond to the existence of repulsive interactions between the pairs of linked nodes, which obligates them to be in separated clusters forming bipartite structures in the network.

From now on we consider, for the sake of simplicity, the case where $\beta = -1$. Then,

$$G_{pq}(\beta = -1) = \left(e^{-\mathbf{A}}\right)_{pq} = \sum_{j=1}^{n} \phi_j(p)\phi_j(q) e^{-\lambda_j}. \qquad (6)$$

Now, let us interpret the exponential negative adjacency matrix. First, we expand it in powers of the adjacency matrix,

$$e^{-A} = I - A + \frac{A^2}{2!} - \frac{A^3}{3!} + \cdots, \tag{7}$$

which can be expressed in terms of the hyperbolic functions as,

$$e^{-A} = \cosh(A) - \sinh(A). \tag{8}$$

The term $[\cosh(A)]_{pq}$ represents the weighted sum of the number of walks of even length connecting nodes $p$ and $q$ in the network. Similarly, $[\sinh(A)]_{pq}$ represents the weighted sum of the number of walks of odd length connecting nodes $p$ and $q$.

Let us consider a bipartite graph and let $p$ and $q$ be nodes which are in two different disjoint sets of the graph. Then, there are no walks of even length between $p$ and $q$ in the graph and

$$G_{pq}(\beta = -1) = \left[-\sinh(A)\right]_{pq} < 0. \tag{9}$$

However, if $p$ and $q$ are nodes in the same disjoint set, then there is no walk of odd length connecting them due to the lack of odd cycles in the bipartite graph, which makes

$$G_{pq}(\beta = -1) = \left[\cosh(A)\right]_{pq} > 0. \tag{10}$$

The above argument shows that, in general, the sign of the communicability at a negative temperature, $G_{pq}(\beta = -1) = \left(e^{-A}\right)_{pq}$, gives an indication as to how the nodes can be separated into disjoint sets.

**Strategy for detecting quasi-bipartite clusters**

The relations (9) and (10) hold when the graph is bipartite. However, the arguments concerning odd versus even length walks carry through to the case of quasi-bipartite subgraphs. A bipartite subgraph (bipartite cluster) is a subset of nodes of the graph that can be divided into two disjoint sets. In a quasi-bipartite subgraph we have a subset of nodes that can be divided into two *almost disjoint* sets. The term "almost disjoint" means that most of the links in the subgraph are inter-set links but there are very few intra-set links. In more formal terms we can define a quasi-bipartite cluster of nodes as follows.

**Definition 1:** Let $C \subseteq V$ be a cluster of nodes in the network. Then, $C$ is a quasi-bipartite cluster if, and only if, $[\cosh(A)]_{pq} > [\sinh(A)]_{pq} \, \forall p, q \in C$.

Our strategy for detecting quasi-bipartite clusters in complex networks is as follows. First we start by calculating $\exp(-\mathbf{A})$, whose $(p,q)$-entry gives the communicability at negative temperature between the nodes $p$ and $q$ in the network. We recall that the positive entries of this matrix, i.e., $G_{pq} > 0$, correspond to pairs of nodes which are in the same quasi-bipartite cluster. The matrix $\exp(-\mathbf{A})$ can be represented as a signed complete graph in which every link connecting a pair of nodes has positive ($G_{pq} > 0$) or negative ($G_{pq} < 0$) signs. We recall that a signed graph is a graph whose edges are labelled by signs. In a signed graph a positive clique is a maximal set of mutually adjacent vertices in which every pair of nodes is linked by a positive edge. That is, a positive subgraph $B$ is said to be a positive clique if $G_{pq} > 0 \; \forall p, q \in B$. Then a *quasi-bipartite cluster* in the network is a positive clique. A clique is a maximal complete subgraph and a complete subgraph is a part of a graph in which all nodes are connected to each other.

In order to account for the inter- to intra-set proportion of links in the detection of quasi-bipartite clusters we introduce a Heaviside step function:

$$\Theta(x) = \begin{cases} 1 \text{ if } x > 0 \\ 0 \text{ if } x \leq 0 \end{cases}.$$

**Definition 2**. The *node-repulsion graph* is a graph whose adjacency matrix is given by $\Theta[\exp(-A)]$, which results from the elementwise application of the function $\Theta(x)$ to the matrix $\exp(-A)$. A pair of nodes $p$ and $q$ in the node-repulsion graph $\Theta[\exp(-A)]$ is connected if, and only if, they have $G_{pq} > 0$.

Now, suppose that there is a link between the nodes $p$ and $q$ and there are also links between them and a third node $r$. This means that $G_{pq} > 0$, $G_{pr} > 0$ and $G_{qr} > 0$. Consequently, the three nodes form a positive subgraph $B$. As we want to detect the largest subset of nodes connected to this triple we have to search for the nodes $s$ for which $G_{is} > 0$ $\forall i \in B$. Using the *node-repulsion graph,* this search is reduced to finding the cliques in a simple graph, $\Theta[\exp(-A)]$. These cliques correspond to the quasi-bipartite clusters of the network.

Finding the cliques in a graph is a classical problem in combinatorial optimization, which has found applications in diverse areas [23]. Here we use a well-known algorithm due to Bron and Kerbosch [24], which is a depth-first search for generating all cliques in a graph. This algorithm consumes a time per clique which is almost independent of the graph size for random graphs and for the Moon-Moser graphs of $n$ vertices the total time is proportional to $(3.14)^{n/3}$. The Moon-Moser graphs have the largest number of maximal cliques possible among all *n*-vertex graphs regardless of the number of edges in the graph [25].

Our new algorithm differs from those in [9] and [10] in a number of ways. Fundamentally, our aim is different. Rather than quantifying the overall bipartivity of the network, or of individual nodes or edges, we are looking for communities that share in a bipartite substructure. Moreover, we allow several such substructures to be present. In contrast to [9], the new algorithm takes account of both odd and even length walks, avoids the need for a cutoff parameter by considering walks of all possible lengths, and does not require a complex energy landscape to be searched by a heuristic discrete optimization algorithm. The new approach differs from [10] in that the *difference* between odd and even length walks is used, and we consider walks between distinct pairs of nodes rather than closed walks, allowing communities to be detected.

**Computational tests**

In order to test our strategy to find quasi-bipartite clusters in complex networks, we start with a small quasi-bipartite graph. The graph was formed by placing 12 nodes into two disjoint sets of 6 nodes each (a bipartite graph) and then connecting at random some of the nodes in each disjoint set, which creates a quasi-bipartite structure. The graph has average degree $\langle k \rangle = 6.67$. We apply the Bron and Kerbosch algorithm to the 0,1-adjacency matrix of the the node-repulsion graph, $\Theta[\exp(-A)]$, to identify the cliques, which correspond to the quasi-bipartite clusters in the original graph.

In Fig. 1A, we illustrate the graph in such a way that the quasi-bipartite structure is not visually apparent. This would be the case with a typical graph drawing algorithm. The node-repulsion graph $\Theta[\exp(-A)]$, which consists of two isolated components, is illustrated in Fig. 1B. Each of these components is formed by a complete graph of 6 nodes. Then, it is evident that the BK algorithm identifies these two cliques as the only ones existing in the node-repulsion graph $\Theta[\exp(-A)]$, which indeed correspond to the two quasi-bipartite clusters of the graph, as illustrated in Fig. 1C. The two almost disjoint sets are represented in two lines of nodes where the inter-cluster links are represented in gray and the intra-cluster ones in black. The intra-cluster links make the graph quasi-bipartite rather than perfectly bipartite.

**Insert Fig. 1 about here.**

Next, we create two new graphs using a similar approach. The first is a quasi-tripartite graph and the second is a quasi-tetrapartite graph. The application of the current approach clearly divides the first graph into three quasi-partite clusters and the second one into four. The graphs and their partitions are illustrated in Figs. 2 and 3, respectively.

<div align="center">**Insert Figs. 2 and 3 about here.**</div>

**Multipartite Structure in Real-World Networks**

As a proof of concept we first select a network which we already know is bipartite. It is the network of heterosexual contacts obtained empirically at the Cadham Provincial Laboratory during 6 months between November 1997 and May 1998 [26]. This network, consisting of 82 nodes and 84 connections, was studied by Lind et al. [27] where illustrations and details can be found. Our current approach clearly identifies the two bipartite clusters, one consisting of 47 nodes and the other of 35 (results not shown). The node-repulsion graph clearly identified the two isolated components.

As a second example, we studied the food web of Canton Creek, which consists primarily of invertebrates and algae in a tributary, surrounded by pasture, of the Taieri River in the South Island of New Zealand [28]. This network consists of 108 nodes (species) and 707 links (trophic relations). Using our current approach, we find that this network can be divided into two almost-bipartite clusters, one having 66 nodes and the other 42. Only 20 links connect nodes in the same clusters, 13 of them connect nodes in the set containing 66 nodes and the other 7 connect nodes in the set of 42 nodes. Thus 97.2% of links are connections between the two almost-bipartite clusters and only 2.8% links are intracluster connections. In Fig. 4, we illustrate the network and its quasi-bipartite clusters as found in the current work. Other food webs (see [29] and the references therein), like that of the pelagic species from the largest of a set of 50 New York Adirondack lake food webs (Bridge Brook), a marine ecosystem on the northeast US shelf (Shelf), invertebrates in an English pond (Skipwith) and a food web like Canton Creek but in native tussock habitat (Stony stream) are also formed by two main quasi-bipartite clusters with no overlap between them. However, there are other food webs with a larger number of quasi-bipartite clusters with large overlap among them.

One example is the network formed by birds and predators and arthropod prey of *Anolis* lizards on the island of St. Martin (see [29] and the references therein), located in the northern Lesser Antilles (StMartins), which has 116 quasi-partite clusters.

**Insert Fig. 4 about here.**

The next example corresponds to the protein-protein interaction network (PIN) of the Kaposi sarcoma-associated herpesvirus (KSHV) [30]. KSHV is a member of the $\gamma$-herpesvirus subfamily associated with Kaposi sarcoma and B cell lymphomas. Its PIN was generated by Uetz et al. [30] by testing 12,000 viral protein interactions involving both full-length proteins and protein fragments. From this pool of interactions, Uetz et al. [30] identified 123 nonredundant interacting pairs of proteins, 8 of which were self-interactions. The resulting PIN of KSHV, formed by 50 proteins and 115 interactions, is illustrated in Fig. 5A. Some of the global topological characteristics of this PIN can be found in Uetz. et al. [30].

Using our current approach we identify 34 quasi-partite clusters in the PIN of KSHV. The proteins grouped in every cluster are given in Table 1. The largest clusters are the number 10, 11, 14 and 18 which have 21 proteins each. However, there is a very large overlap among them ranging from 66.6% for the clusters 11 and 18 to 95% for the pairs of clusters (10, 11), (11, 14), (14, 18). There is another group of quasi-bipartite clusters containing a large number of proteins. They are the clusters 28, 29, 30 and 32. They also display very large overlapping among them, ranging from 89.5% to 94.7%. However, these two groups of clusters are completely orthogonal. That is, absolutely no overlapping exists between any of the clusters of the first group (10, 11, 14 and 18) with the clusters in the second group (28, 29, 30 and 32). Then we conclude that the PIN of KSHV can be divided into two disjoint clusters of almost the same size, which contain 78% of the proteins in the PIN. These two clusters are illustrated in the Fig. 5B.

On the other side of the coin there are networks displaying a huge number of small quasi-bipartite clusters. This is the case for those networks lacking a bipartite structure at all but having a multipartite structure. For instance, the neuronal synaptic network of the nematode *C. Elegans* (see [31] and the references therein), which has 280 nodes and 1973 links is formed by 43, 753 quasi-partite clusters. This network has been formerly shown to have a super-homogeneous structure [32], which explains its lack of bipartivity.

**Conclusions**

Given a complex network, the new concept of a node-repulsion graph has intuitive interpretations in terms of (a) a Green's function at negative absolute temperature, and (b) a measure of the discrepancy between the overall number of odd and even walks between pairs of nodes. Moreover, this concept allows for a natural, well-defined quantification of quasi-bipartite clusters that can be investigated with a simple, parameter-free computational algorithm. This new algorithm was able to discover inherent bipartite communities in real data sets, and hence has the potential to unlock hidden patterns at the heart of complex networks.

**Acknowledgements**

EE thanks the support by the Royal Society of Edinburgh and the Edinburgh Mathematical Society (March 2008) and to the IIS, University of Tokyo for a fellowship as Research Visitor during April-June, 2008. DJH was supported by Engineering and Physical Sciences Research Council grant GR/S62383/01

Table 1. Quasi-bipartite clusters found in the protein-protein interaction network of the Kaposi sarcome-related herpes virus (KSHV). The numbers are identification labels for the clusters and the proteins which form the cluster are given in the second column.

| No. | Proteins |
|---|---|
| 1 | K1 K15 K3 K5 K8 K8.1 Orf41 Orf52 Orf54 Orf65 Orf67.5 Orf68 Orf74 Orf56 K12 Orf48 Orf61 |
| 2 | K8.1 Orf41 Orf54 Orf65 Orf67.5 Orf68 K12 Orf48 Orf61 Orf39 Orf47 |
| 3 | Orf41 Orf65 K12 Orf48 Orf61 Orf39 Orf47 Orf2 |
| 4 | Orf48 Orf61 Orf39 Orf47 Orf45 Orf2 Orf28 Orf37 Orf49 |
| 5 | Orf56 Orf48 Orf61 Orf45 |
| 6 | K1 K15 K3 K5 K8 K8.1 Orf23 Orf41 Orf50 Orf52 Orf54 Orf65 Orf67.5 Orf68 Orf74 Orf56 K12 Orf61 |
| 7 | K8.1 Orf41 Orf50 Orf54 Orf65 Orf67.5 Orf68 K12 Orf61 Orf39 Orf47 |
| 8 | Orf61 Orf39 Orf47 Orf45 Orf2 Orf27 Orf28 Orf31 Orf37 Orf49 Orf59 |
| 9 | Orf61 Orf45 Orf2 Orf27 Orf28 Orf31 Orf34 Orf37 Orf49 Orf59 Orf69 |
| 10 | K1 K10 K11 K15 K3 K5 K8 K8.1 Orf23 Orf30 Orf41 Orf50 Orf52 Orf54 Orf58 Orf65 Orf67.5 Orf68 Orf72 Orf74 Orf56 |
| 11 | K1 K10 K10.5 K11 K15 K3 K5 K8 K8.1 Orf23 Orf30 Orf41 Orf50 Orf52 Orf54 Orf58 Orf65 Orf67.5 Orf68 Orf72 Orf74 |
| 12 | K10 K10.5 K11 Orf23 Orf50 Orf72 Orf36 |
| 13 | K10 Orf36 Orf25 |
| 14 | K1 K10.5 K11 K15 K3 K5 K8 K8.1 Orf23 Orf30 Orf41 Orf50 Orf52 Orf54 Orf58 Orf65 Orf67.5 Orf68 Orf72 Orf74 K12 |
| 15 | K10.5 K8.1 Orf41 Orf50 Orf54 Orf65 Orf67.5 Orf68 K12 Orf39 Orf47 |
| 16 | K1 K10.5 K11 K15 K3 K5 K8.1 Orf30 Orf41 Orf52 Orf54 Orf65 Orf67.5 Orf68 Orf74 K12 K9 |
| 17 | K10.5 K8.1 Orf41 Orf54 Orf65 Orf67.5 Orf68 K12 K9 Orf39 Orf47 |
| 18 | K1 K11 K15 K3 K5 K8 K8.1 Orf23 Orf30 Orf41 Orf50 Orf52 Orf54 Orf58 Orf65 Orf67.5 Orf68 Orf72 Orf74 Orf56 K12 |
| 19 | K1 K11 K15 K3 K5 K8.1 Orf30 Orf41 Orf52 Orf54 Orf65 Orf67.5 Orf68 Orf74 Orf56 K12 K9 |
| 20 | K1 K15 K3 K5 K8 K8.1 Orf30 Orf41 Orf52 Orf54 Orf58 Orf65 Orf67.5 Orf68 Orf74 Orf56 K12 Orf48 |
| 21 | K1 K15 K3 K5 K8.1 Orf30 Orf41 Orf52 Orf54 Orf65 Orf67.5 Orf68 Orf74 |
|    | Orf56 K12 K9 Orf48                                                                      |
| 22 | K8.1 Orf41 Orf54 Orf65 Orf67.5 Orf68 K12 K9 Orf48 Orf39 Orf47                           |
| 23 | Orf41 Orf65 K12 K9 Orf48 Orf39 Orf47 Orf2                                               |
| 24 | K9 Orf48 Orf39 Orf47 Orf45 Orf2 Orf37                                                   |
| 25 | Orf56 K9 Orf48 Orf45                                                                    |
| 26 | K9 Orf45 Orf2 Orf37 Orf57                                                               |
| 27 | Orf36 Orf25 K7 Orf27 Orf29b Orf31 Orf34 Orf53 Orf57 Orf59 Orf6 Orf60 Orf62 Orf63 Orf69 Orf9 |
| 28 | Orf25 Orf45 K7 Orf27 Orf28 Orf29b Orf31 Orf34 Orf37 Orf49 Orf53 Orf57 Orf59 Orf6 Orf60 Orf63 Orf69 Orf75 Orf9 |
| 29 | Orf25 K7 Orf27 Orf28 Orf29b Orf31 Orf34 Orf37 Orf49 Orf53 Orf57 Orf59 Orf6 Orf60 Orf62 Orf63 Orf69 Orf75 Orf9 |
| 30 | Orf45 K7 Orf2 Orf27 Orf28 Orf29b Orf31 Orf34 Orf37 Orf49 Orf53 Orf57 Orf59 Orf6 Orf60 Orf63 Orf69 Orf75 Orf9 |
| 31 | Orf39 Orf47 Orf45 K7 Orf2 Orf27 Orf28 Orf31 Orf37 Orf49 Orf53 Orf59 Orf6 Orf60          |
| 32 | 32: K7 Orf2 Orf27 Orf28 Orf29b Orf31 Orf34 Orf37 Orf49 Orf53 Orf57 Orf59 Orf6 Orf60 Orf62 Orf63 Orf69 Orf75 Orf9 |
| 33 | Orf48 Orf45 Orf2 Orf28 Orf37 Orf49 Orf75                                                |
| 34 | Orf56 Orf48 Orf45 Orf75                                                                 |

Fig. 1. A. Quasi-bipartite graph having 12 nodes which was built from a complete bipartite graph with two disjoint sets of 6 nodes each (see main text). B. The node-repulsion graph corresponding to the quasi-bipartite graph shown in Fig. 1A. C. Organization of the nodes of the graph given in Fig. 1A to represent the two quasi-bipartite clusters found by the method developed in this work. The black lines represent the intracluster connections and the gray lines the intercluster links.

Fig. 2. A. Quasi-tripartite graph having 18 nodes which was built from a complete tripartite graph with three disjoint sets of 6 nodes each (see main text). B. Organization of the nodes of the graph given in Fig. 2A to represent the three quasi-tripartite clusters found by the method developed in this work. The black lines represent the intracluster connections and the gray lines the intercluster links.

Fig. 3. A. Quasi-tetrapartite graph having 24 nodes which was built from a complete tetrapartite graph with three disjoint sets of 6 nodes each (see main text). B. Organization of the nodes of the graph given in Fig. 3A to represent the four quasi-tetrapartite clusters found by the method developed in this work. The black lines represent the intracluster connections and the gray lines the intercluster links.

Fig. 4. A. Network representation of the food web of Canton Creek. B. Bipartite structure of this network as found by the method developed in this work. Nodes in each quasi-bipartite cluster are represented by squares and circles. The black lines represent the intracluster connections and the gray lines the intercluster links.

Fig. 5. A. Network representation of the protein-protein interaction network of the Kaposi sacrme-related herpes virus (KSHV). B. Bipartite structure of this network as found by the method developed in this work. Nodes in each quasi-bipartite cluster are represented by squares and circles. Triangles correspond to nodes not in these two quasi-bipartite groups (see text for explanation). The black lines represent the intracluster connections and the gray lines the intercluster links

**Figure 1**

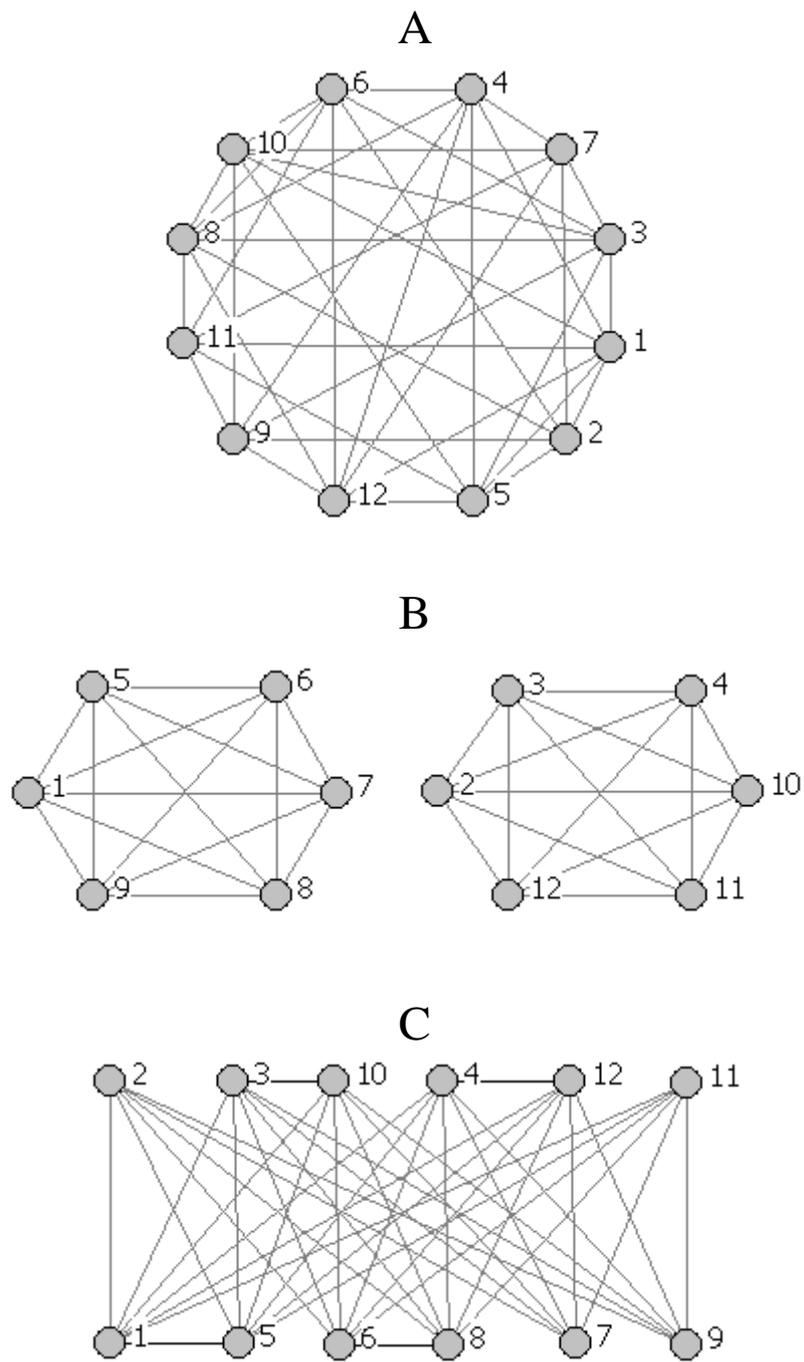

**Figure 2**

A

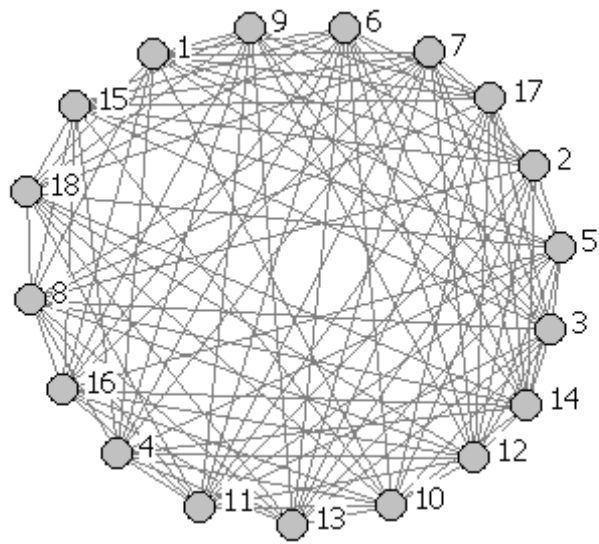

B

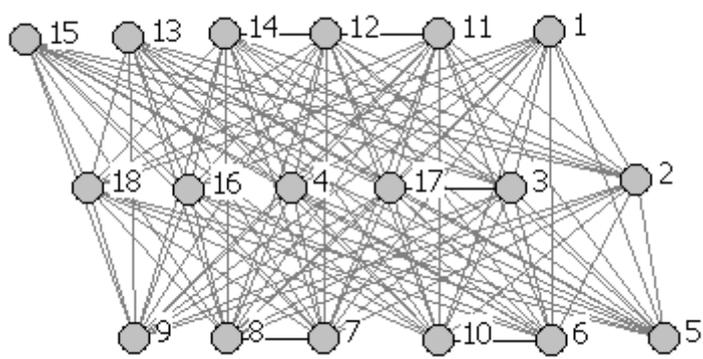

**Figure 3**

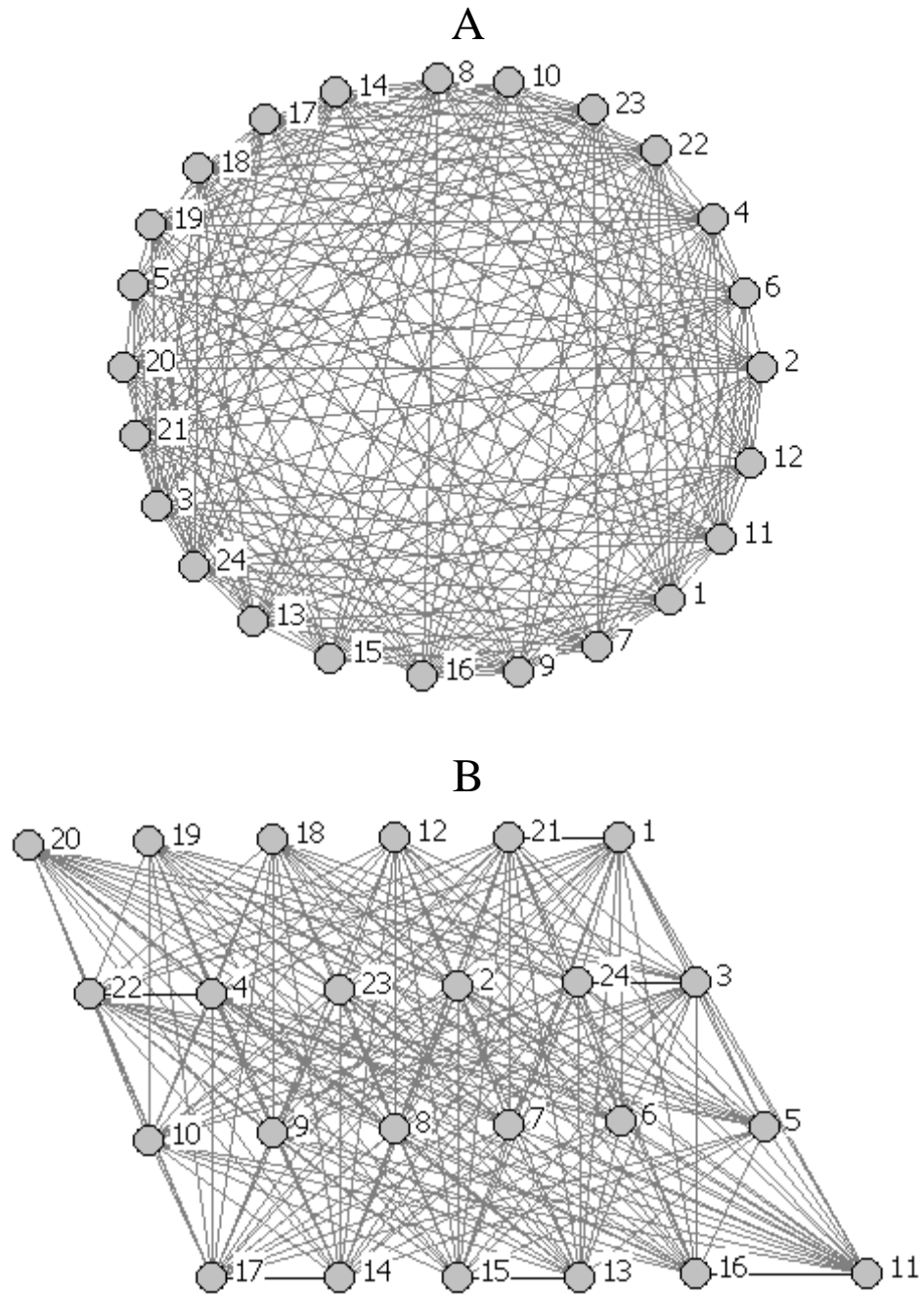

**Figure 4**

A

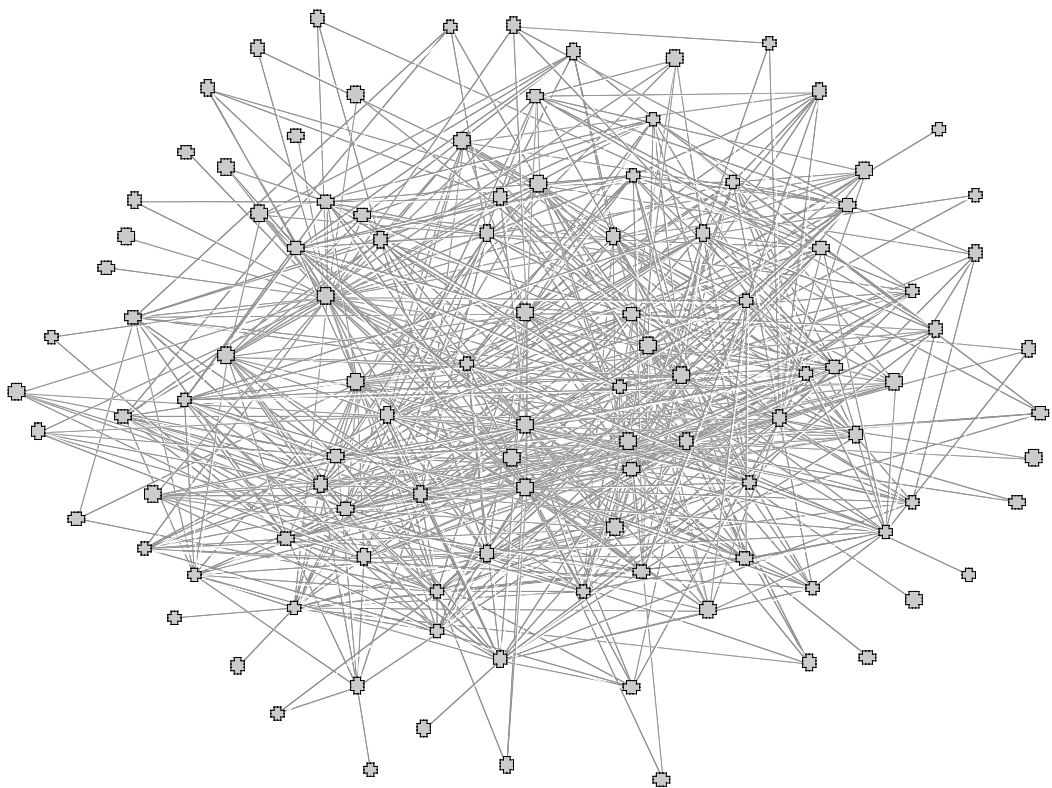

B

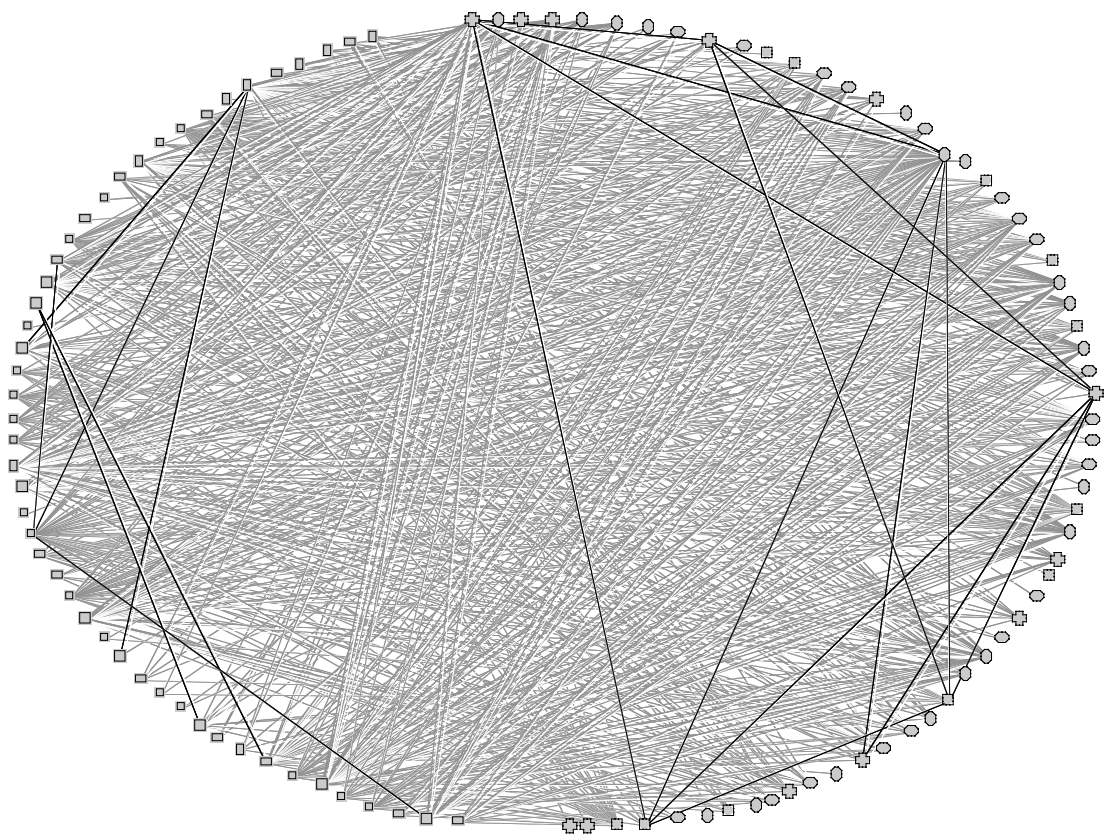

**Figure 5**

A

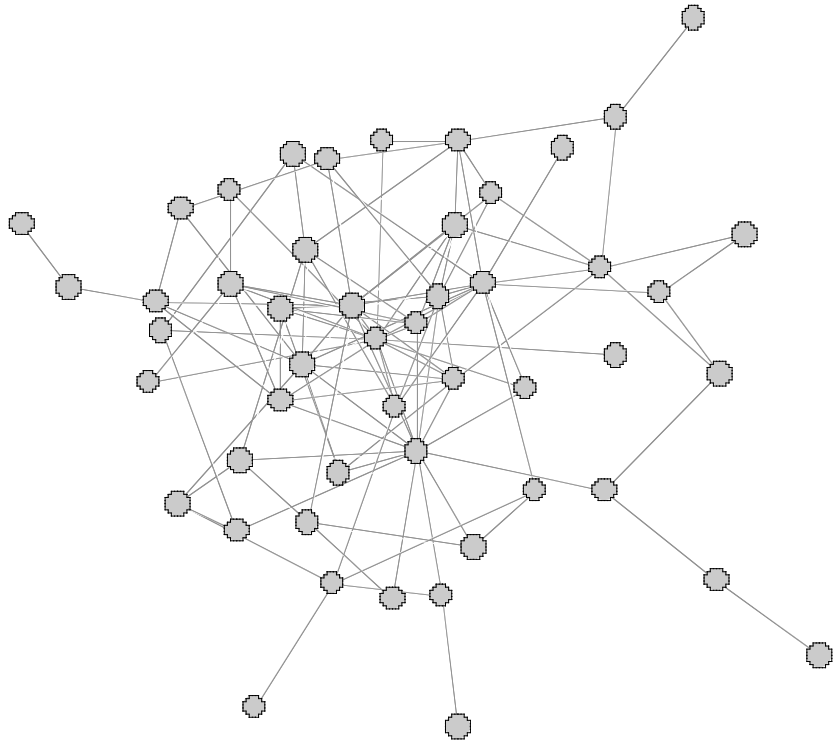

B

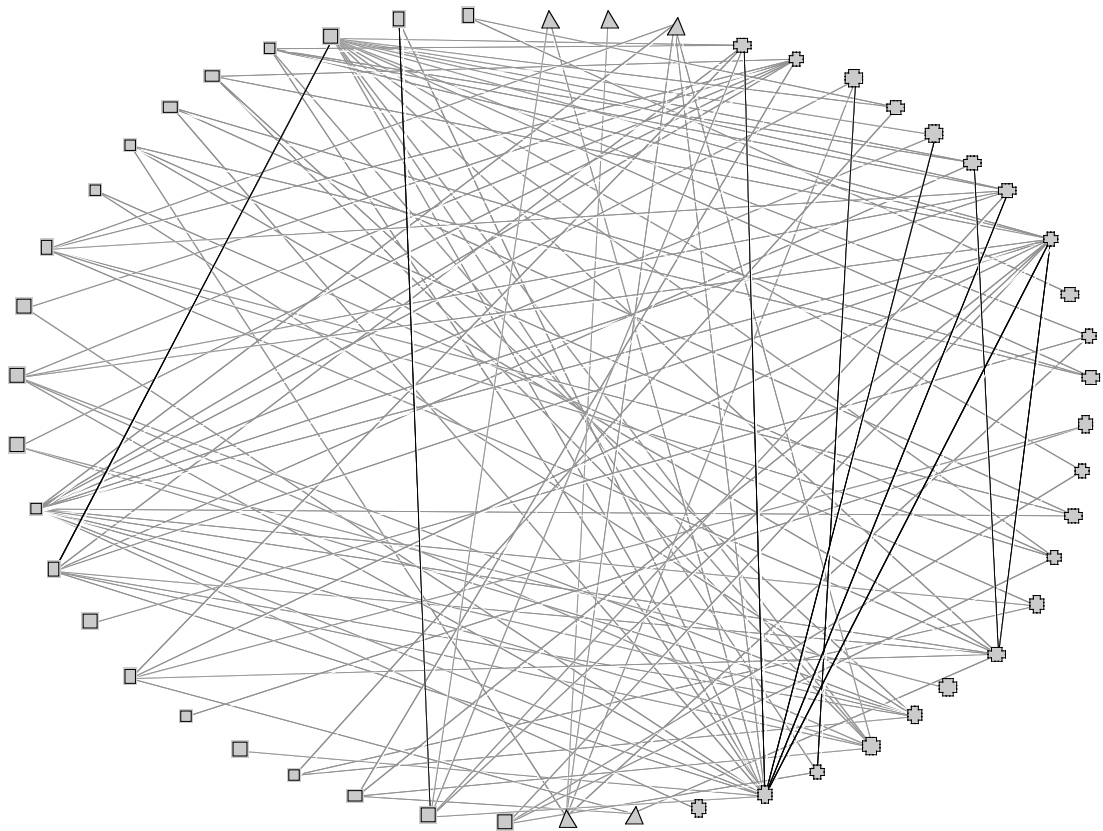